\documentclass[doublecol]{epl2} 
\usepackage{epstopdf}

\title{Non-extensive processes associated with heating of the Galactic disc}
\shorttitle{Non-extensive processes associated with heating of the Galactic disc} 

\author{C. A. P. Viana\inst{1} and D. B. de Freitas\inst{2,3}}
\shortauthor{Viana and de Freitas}

\institute{
\inst{1} Departamento de F\'{\i}sica, Universidade Federal do Rio Grande do Norte, 59072-970 Natal, RN, Brazil\\
\inst{2} Departamento de F\'{\i}sica, Universidade Federal do Cear\'a, Caixa Postal 6030, Campus do Pici, 60455-900 Fortaleza, Cear\'a, Brasil\\
\inst{3} INAF-Osservatorio Astrofisico di Catania, Via S. Sofia 78, I-95123, Catania, Italy}

\abstract{We analyse the mechanisms ruling galactic disc heating through the dynamics of space velocities $U$, $V$ and $W$, extracted from the Geneva-Copenhagen catalogue. To do this, we use a model based on non-extensive statistical mechanics, where we derive the probability distribution functions that quantify the non-Gaussian effects. Furthermore, we find that the deviation $q-1$ at a given stellar age follows non-random behaviour. As a result, the $q$-index behaviour indicates that the vertical component $W$, perpendicular to the Galactic plane, does not ``heat up'' at random, which is in disagreement with previous works that attributed the evolution of $W$ to randomness. Finally, our results bring a new perspective to this matter and open the way for studying Galactic kinematic components through the eyes of more robust statistical models that consider non-Gaussian effects.}

\begin{document}

\maketitle

\section{Introduction}
The components of Galactic space velocity in the Galactic disc are usually specified by the letters $U$, $V$, and $W$ (in km/s), where $U$ denotes the radial component towards the Galactic centre, $V$ represents the tangential component in the direction of Galactic rotation and $W$ designates the vertical component that is perpendicular to the Galactic plane towards the North Galactic Pole (NGP; see Ref \cite{john} for further details).

The observed space velocity distributions of these components are fundamental to deducing how the disc has evolved chemically and dynamically \cite{Stromgren,Stromgren2,Nordstrom2,Chiba}. To this end, it is necessary to have a catalogue that provides an unprecedented sample to investigate this important and current issue of galactic dynamics \cite{Gilmore,newton}. In this context, F and G dwarfs extracted from the Geneva-Copenhagen Survey (GCS) allow more accurate determinations of the space distribution of disc stars in the corresponding age interval. In particular, a deeper investigation of kinematic heating of discs in their space velocities must take into account the stellar age, which is a crucial element for understanding in terms of the evolution of the disc itself \cite{Quinn,Nordstron}.

According to Nordström {\it et al.} \cite{Nordstron}, at least four mechanisms have been proposed to explain the heating of the Galactic disc: i) fast perturbers from the halo, such as massive black holes; ii) slow perturbers in the disc, such as giant molecular clouds; iii) large scale perturbations of the disc caused by spiral arms \cite{deSimone}; and iv) heating caused by in-falling satellite galaxies. Among these options, De Simone {\it et al.} 's proposal is more attractive and plausible, as highlighted by Nordströn {\it et al.} \cite{Nordstron}. As mentioned by De Simone {\it et al.} \cite{deSimone}, the effects of stochastic, transient spiral wave structures can produce exponents in the observed range. 

On the other hand, Holmberg {\it et al.} \cite{Holmberg, Holmberg2} illustrated significant structures in the $U$ and $V$ velocity distributions that persist over a wide range of ages up to at least the Solar age (see also \cite{monari,Schonrich2}), indicating a behaviour similar to non-random processes. In contrast, these authors cited that the $W$ velocities show no deviation from a random distribution in any age range, suggesting that different heating mechanisms are acting in the plane of the disc and perpendicular to it \cite{Casagrande,Lacey,Mignard,deSouza}. 

In the present study, we explore the behaviour of the space velocity distribution in the solar neighbourhood, based on non-extensive statistical mechanics. Our investigation is based on the hypothesis of non-random processes associated with heating the Galactic disc \cite{deSimone}. We find that the non-random process can be explained by the profile of $q$-index as a function of stellar age \cite{deFreitas2,deFreitas}.

According to Nordström {\it et al.} \cite{Nordstron} and Holmberg {\it et al.} \cite{Holmberg, Holmberg2}, the behaviour of velocity $W$ is associated with random (pure) heating. On the other hand, the mechanisms that act on the $U-V$ plane cause a non-random diffusion. These authors suggest that the $U$ and $V$ components of velocity do not follow a purely Gaussian profile, therefore only the $W$ component would have the behaviour described by a Gaussian distribution. Recently, de Freitas and de Medeiros \cite{deFreitas2}, using the GCS catalogue, showed that the best fit for the radial velocity distribution segregated by the age of F- and G-type field stars is driven by a $q$-Gaussian distribution.

A detailed description of non-extensive frameworks and their properties are shown in Section 2. In Section 3, we describe the sample and how we selected stars and physical stellar parameters for the present analyses. Section 4 brings the main results and discussions of the present analysis. Finally, concluding remarks are presented in the last section.

\begin{figure*}[htb]
\centering
\includegraphics [scale = 0.60] {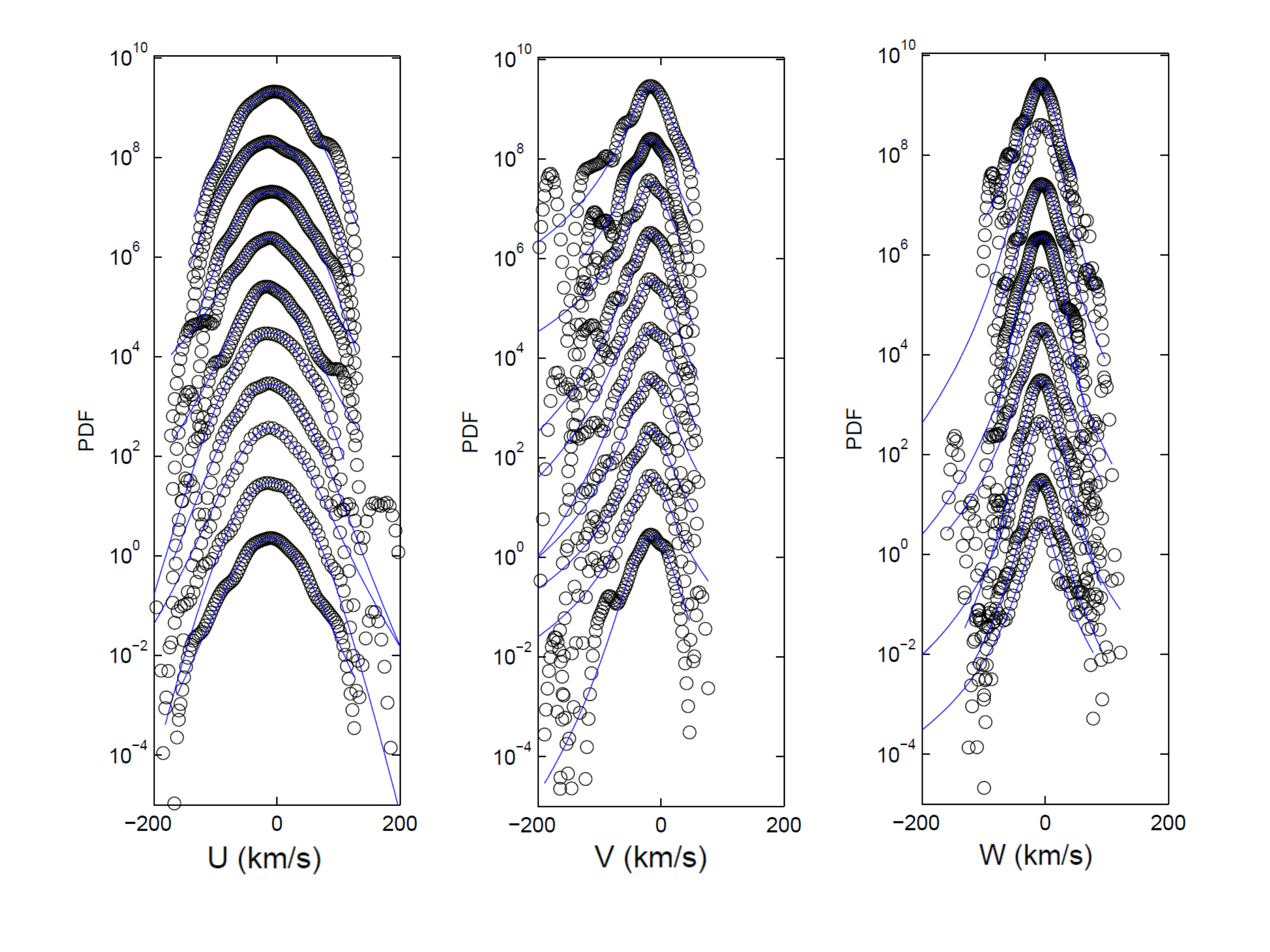}
\caption{Semi-log plot of the PDF (dots)  and $q$-Gaussian (blue curves) distribution of space velocities, $U$, $V$ and $W$, for G-type stars segregated by age from 1 to 10 Gyr (from bottom to top, respectively). The distribution functions are shifted up by a factor of 10 each, for the sake of clarity. As an example, the Sun is classified as a G-type star.}
\label{fig0}
\end{figure*} 

\begin{figure*}[htb]
\centering
\includegraphics [scale = 0.58] {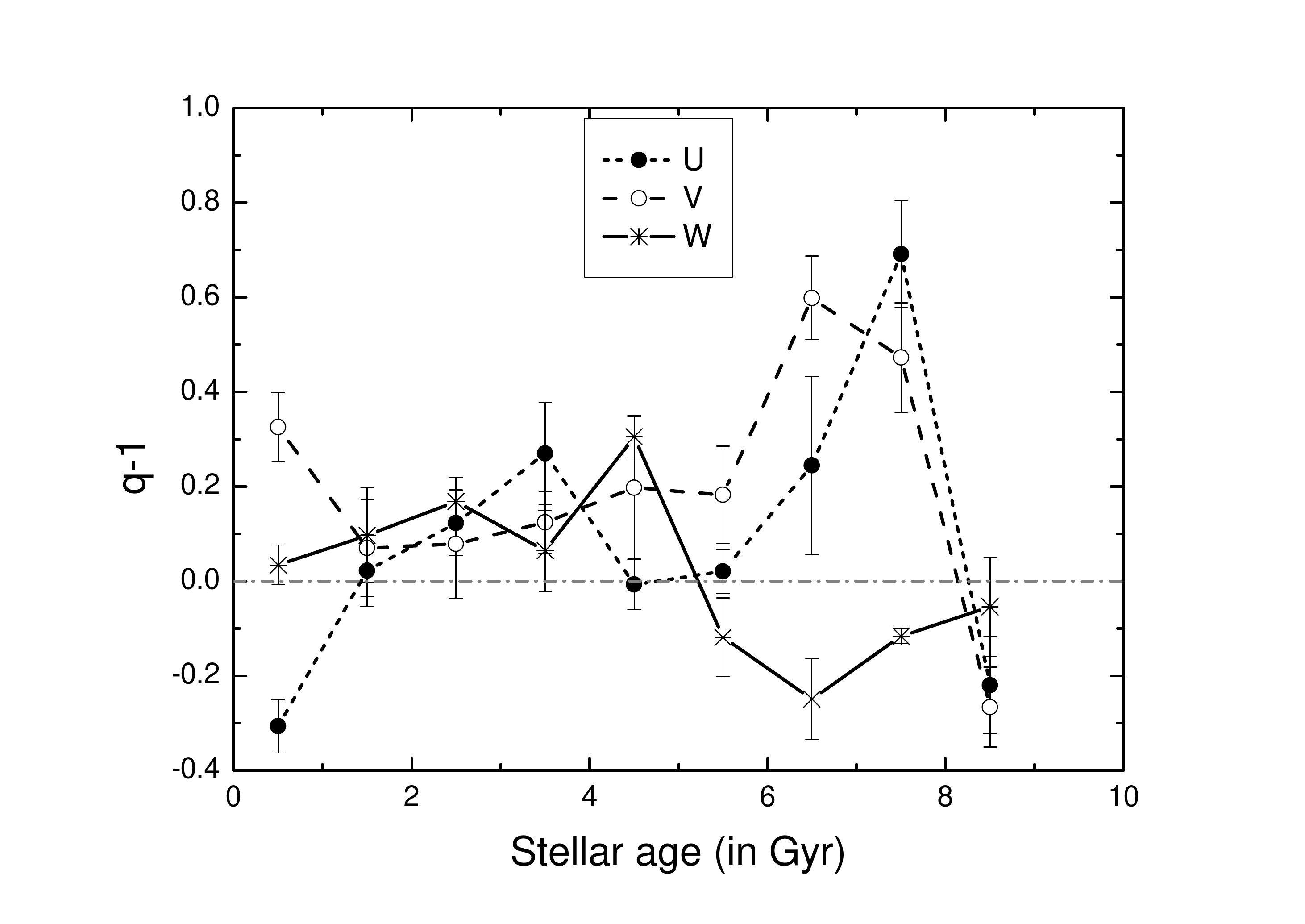}
\caption{Gaussian deviation ($q - 1$) of the distribution of the $U$, $V$ and $W$ components of F-type stars. The error bars are bootstrapped 95\% confidence intervals. The dashed-dot line indicates the gaussianity.}
\label{fig1}
\end{figure*}

\begin{figure*}[htb]
\centering
\includegraphics [scale = 0.58] {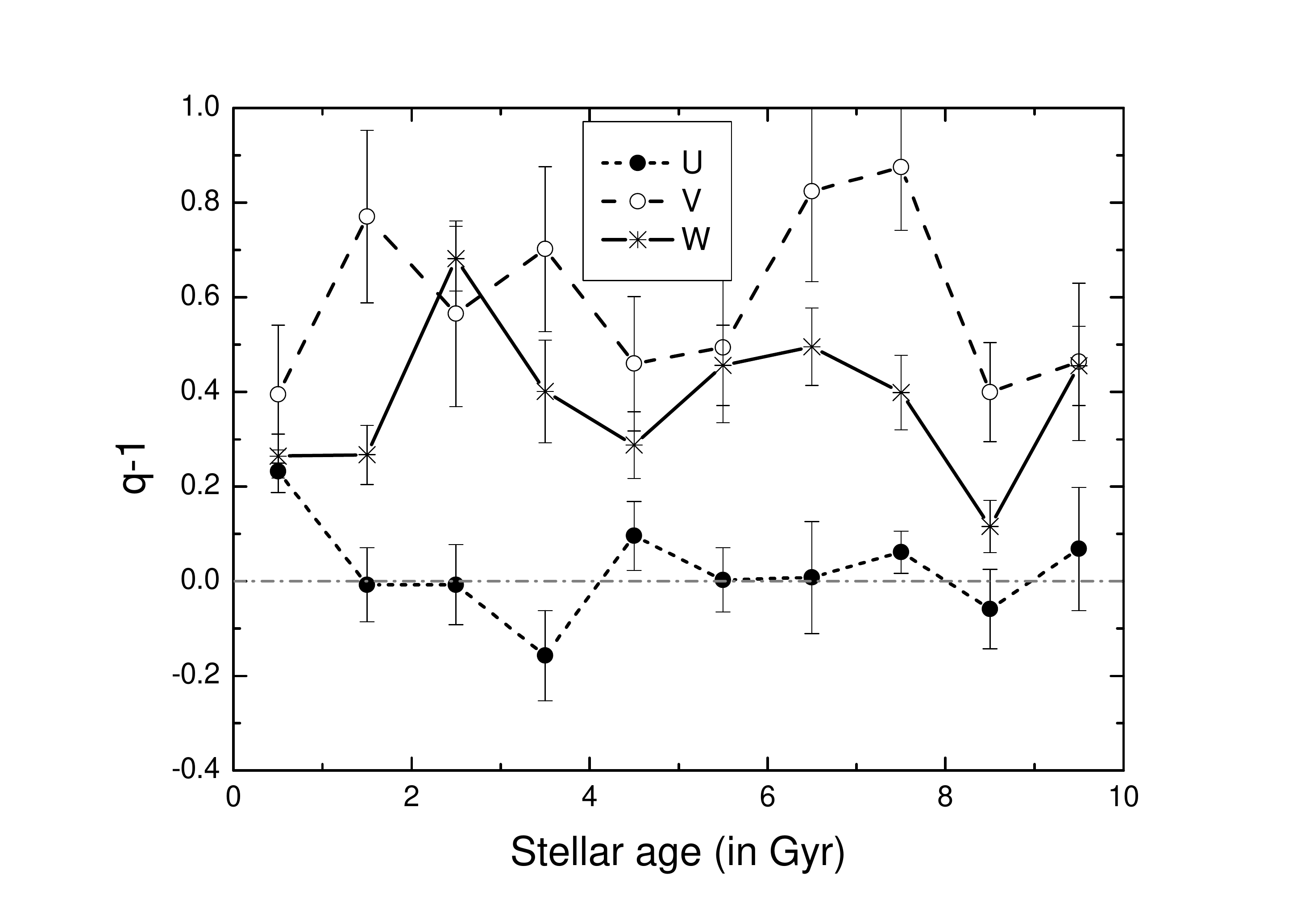}
\caption{Idem Figure \ref{fig1} for G-type stars.}
\label{fig2}
\end{figure*}

\section{Nonextensive formalism}
The prototype of entropy that we are considering is Boltzmann-Gibbs-Shannon (BGS) entropy. To mention only the most familiar statistics, this entropy has been generalised by other "entropy-like" indexes, which emerge from approaches such as Kolmogorov-Sinai entropy (dynamical systems)\cite{kol}, Rényi entropy (information theory)\cite{ren}, Kaniadakis entropy (relativistic kinetic theory)\cite{kani} and Tsallis entropy (statistical physics)\cite{tsa}. Roughly speaking, the essence of these new entropic forms is to recover BGS entropy from Khinchin-Shannon axioms (for futher details see Ref. \cite{amigo}). In particular, physical systems with long range interactions defy the fourth axiom, also called the additivity axiom, i.e., $S(A+B)=S(A)+S(B)$. To do this, a non-additive entropy uses an interation term between the sub-systems $A$ and $B$. In this context, the non-additivity poses the question of what mathematical properties have the ``generalised entropies'' satisfying this axiom. Our choice of the Tsallis entropy form is justified by its unique properties under those axioms, in particular the additivity axiom. As an example, gravitational interaction is an interesting case of long range interactions \cite{gell}.

de Freitas and de Medeiros \cite{deFreitas2}, using the same dataset as the present paper, showed that the high values of $q$ extracted from radial velocity distributions reveal the effects of long-range interactions consistent with the non-extensive central limit theorem ($q$-CLT). Because kinematics and physical properties of the space velocities of stars are defined by gravitational interaction, we choose the most appropriate non-additive entropic form with a wide range of tested systems. As an example, Kolmogorov-Sinai, Kaniadakis and Rényi entropy are additive and, therefore, are at a disadvantage compared to Tsallis entropy, at least for the present case.

In this context, Tsallis' $q$-entropy, $S_{q}$, \cite{Tsallis1,Tsallis3,Tsallis4} considers that a system composed by two correlated sub-systems, $A$ and $B$, has entropy defined by:

\begin{eqnarray}
\label{eq1}
S_{q}^{(A+B)} = S_{q}^{(A)} + S_{q}^{(B)} + \frac{(1-q)}{k}S_{q}^{(A)} S_{q}^{(B)},
\end{eqnarray}
where $q$ is the entropic index that characterises generalisation and $k$ is Boltzmann's constant. When $q\rightarrow1$, the extensive entropy is recovered. The values of $q$-entropic indexes are derived from probability distribution functions (PDF) named by $q$-Gaussians, $p_{q}(x)$. These functions are obtained from the variational problem using the continuous version for the non-extensive entropy given by Eq. (\ref{eq1}) and, therefore, defined by:
\begin{eqnarray}
\label{eq2}
p_{q}(x) = A_{q}[1 +(1-q)B_{q}x^{2}]^{1/(1-q)}, 
\end{eqnarray}
where $x$ denotes the velocities $U$, $V$ and $W$. For the parameter $A_{q}$, there are two conditions:

(i) $q<1$, 
\begin{eqnarray}
\label{eq3}
A_{q}=\frac{\Gamma \left[\frac{5-3q}{2-2q}\right] }{\Gamma \left[ \frac{2-q}{1-q}\right]}\sqrt{\frac{1-q}{\pi}B_{q}}
\end{eqnarray}
and (ii) $q>1$,
\begin{eqnarray}
\label{eq4}
A_{q} = \frac{\Gamma \left[\frac{1}{q-1}\right] }{\Gamma \left[ \frac{3-q}{2q-2}\right]}\sqrt{\frac{q-1}{\pi}B_{q}}.
\end{eqnarray}

The value of $B_{q}$ is a function of variance $\sigma_{q}$ and is given by: 
\begin{eqnarray}
\label{eq5}
B_{q} = [(3-q)\sigma_{q}^{-2}]^{-1}.
\end{eqnarray}
For further details, see Ref. \cite{Tsallis1,Tsallis3,deFreitas}. In the present work, the behaviour of parameter $B_{q}$ as a function of stellar age was not relevant and we discarded it from the results.

\begin{table}
\par
\centering
\par
\caption{\label{tab1} Best values of bootstrap $q$-median and their errors using the bootstrapped 95\% confidence interval ($ci$) of distributions of the space velocities $U$, $V$ and $W$ for F-type stars.
}
\begin{tabular}{cccccccc}
\hline \hline\noalign{\smallskip}

        \textbf{Age}
         
        & \multicolumn{5}{c}{\textbf{F- type stars}} \\ (Gyr) & ${\it q}_{U}$ & $ci_{U}$ & ${\it q}_{V}$ & $ci_{V}$ & ${\it q}_{W}$ & $ci_{W}$  \\
                
        \noalign{\smallskip}
        \hline\noalign{\smallskip}
0 $-$ 1	&	0.69	&	0.06	&	1.33	&	0.33	&	1.03	&	0.04	\\
1 $-$ 2	&	1.02	&	0.08	&	1.07	&	0.07	&	1.10	&	0.10	\\
2 $-$ 3	&	1.12	&	0.07	&	1.08	&	0.08	&	1.17	&	0.05	\\
3 $-$ 4	&	1.27	&	0.11	&	1.12	&	0.12	&	1.06	&	0.09	\\
4 $-$ 5	&	0.99	&	0.05	&	1.20	&	0.20	&	1.31	&	0.05	\\
5 $-$ 6	&	1.02	&	0.05	&	1.18	&	0.18	&	0.88	&	0.08	\\
6 $-$ 7	&	1.24	&	0.19	&	1.60	&	0.60	&	0.75	&	0.09	\\
7 $-$ 8	&	1.69	&	0.11	&	1.47	&	0.47	&	0.88	&	0.02	\\
8 $-$ 9	&	0.78	&	0.10	&	0.73	&	-0.27	&	0.95	&	0.10	\\
9 $-$ 10	&	-	&	-	&	-	&	-	&	-	&	-	\\

\hline
\end{tabular}
\end{table}

\begin{table}
\par
\centering
\par
\caption{\label{tab2} Idem Table 1 for G-type stars.
        }
\begin{tabular}{cccccccc}
\hline \hline\noalign{\smallskip}

        \textbf{Age}
         
        & \multicolumn{5}{c}{\textbf{G- type stars}} \\ (Gyr) & ${\it q}_{U}$ & $ci_{U}$ & ${\it q}_{V}$ & $ci_{V}$ & ${\it q}_{W}$ & $ci_{W}$  \\

        \noalign{\smallskip}
        \hline\noalign{\smallskip}

0 $-$ 1	&	1.23	&	0.05	&	1.40	&	0.15	&	1.26	&	0.05	\\
1 $-$ 2	&	0.99	&	0.08	&	1.77	&	0.18	&	1.27	&	0.06	\\
2 $-$ 3	&	0.99	&	0.08	&	1.56	&	0.20	&	1.68	&	0.07	\\
3 $-$ 4	&	0.84	&	0.10	&	1.70	&	0.17	&	1.40	&	0.11	\\
4 $-$ 5	&	1.10	&	0.07	&	1.46	&	0.14	&	1.29	&	0.07	\\
5 $-$ 6	&	1.00	&	0.07	&	1.49	&	0.16	&	1.46	&	0.08	\\
6 $-$ 7	&	1.01	&	0.12	&	1.82	&	0.19	&	1.50	&	0.08	\\
7 $-$ 8	&	1.06	&	0.04	&	1.87	&	0.13	&	1.40	&	0.08	\\
8 $-$ 9	&	0.94	&	0.08	&	1.40	&	0.10	&	1.12	&	0.05	\\
9 $-$ 10	&	1.07	&	0.13	&	1.46	&	0.17	&	1.46	&	0.08	\\

\hline
\end{tabular}
\end{table}

\section{Working sample}
Obtaining the heliocentric space velocities needed to exploit the present scientific case requires astrometric data that can be extracted from different missions. Among them, GAIA and Hipparcos space missions are relevant, but they do not have a sufficient range of stellar parameters, such as projected rotational velocity $v\sin i$, age and spectral type, among others, which are indispensable for a more accurate evaluation of the results (see also \cite{udry,gaia}). Undoubtedly, GAIA is far superior to the Hipparcos mission. The ESA pioneer mission GAIA launched in 2013 provides an unprecedented quantitative leap in the study of stellar kinematics, providing more than 1 billions targets. However, the GAIA multi-epoch photometric data present much larger radial-velocity errors. For bright stars, GAIA offers an accuracy of 1--2 km $s^{-1}$ in radial velocity, whereas for faint stars, which will be mostly distant stars and of interest as tracers of Galactic dynamics, an accuracy of 5--10 km $s^{-1}$ is obtained, therefore affecting the $U$ component of space velocity \cite{Nordstron,gaia}. Among the catalogs published in the literature, the GCS supports these requirements. The GCS data are still useful even in the face of GAIA data, not only for studying our Galaxy, but also for accurate radial velocity for all targets to a mean propagated error in the space velocity less than 0.7 km $s^{-1}$, as well as a range of stellar parameters useful for analysing results.

The sample used in this work is composed of F- and G- type field dwarf stars located in the solar neighbourhood, complete in magnitude and volume of $\sim 40$ pc, taken from the GCS catalogue, which was published by Nordström {\it et al.} \cite{Nordstron} and later revised by Holmberg {\it et al.} \cite{Holmberg}, \cite{Holmberg2} and Casagrande {\it et al.} \cite{Casagrande}. The catalogue contains data on the age, metallicity, mass, projected rotation velocity ($v\sin i$) and kinematic properties for about 14000 stars in the solar neighbourhood. The space velocity triplets ($U, V, W$) in the GCS catalogue were calculated using the data of distance, proper movement and the average radial velocity; these observed parameters were obtained through CORAVEL and CfA instruments \cite{Nordstron}. To guarantee better precision in the results, Holmberg {\it et al.} \cite{Holmberg2} surveyed stars that had errors in age below 25\% ($\sim$ 2600 stars). In this context, our study is based on well-defined age and mass (spectral type), which are fundamental criteria to investigate a possible relationship between the $U$, $V$ and $W$ components and the entropic index $q$. In addition, we also take into account any possible contamination by binary systems or giants. A more detailed description of the criteria that define the sample is shown below.

Most of the previous studies on this subject used ages estimated from chromospheric activity. However, this procedure is not valid for stars of the Sun's age because chromospheric emission essentially vanishes at this stage. Thus, theoretical isochrone ages are the best choice, although some important discrepancies persist, particularly for old low-mass stars. The ages given in the GCS were then calculated based on the stellar isochrone procedure \cite{Nordstron}, using the Bayesian computational technique of Jorgensen and Lindegren \cite{jorgensen05}. Holmberg {\it et al.} \cite{Holmberg} conducted a review of the data considering new temperature and metallicity calibrations. According to Nordström {\it et al.}\cite{Nordstron} and \cite{Holmberg}, errors in age estimations are below 50$\%$ for 81$\%$ of the presumably single stars in the original GCS sample.

Accurate mass measurements are crucial when investigating the evolutionary history of stars in the sample. Nordström {\it et al.}\cite{Nordstron} estimated stellar masses via theoretical isochrone analysis, using an M-function to describe the probability distribution of model masses from the Padova model for an observed star \cite{girardi2000}, with individual errors averaging about 0.05 M$_{\odot}$. The entire sample consists of stars whose masses range from 0.65 to 2.43 $M_{\odot}$.

In the GCS catalogue, stars with photometric distances that deviate more than the $3\sigma$ from the Hipparcos distances are flagged as suspected binaries or giants. According to Sharma {\it et al.} \cite{sharma11}, we can use the following selection function to mimic the selective avoidance of giants: $M_{V}<10(b-y)-3$, where $M_{V}$ is the absolute magnitude and $b-y$ denotes a colour in the Stromgren $uvby\beta$ system.

Based on the criteria and cuts mentioned above, we obtained a final working sample of 6704 single stars. Our final sample is composed of 3880 F- and 2824 G-type stars, with ages limited to 10 Gyr (the upper limit for ages was defined by the star's lifetime on the main-sequence and the age of the Galactic disc) and masses defined in the range of $0.65<M/M_{\odot}<2.5$.

\section{Results and discussion}
The result of $q$-index as a function of stellar age intervals is presented in Tables \ref{tab1} and \ref{tab2}. These values were derived from PDFs of space velocities. Fig. \ref{fig0} is shown here to illustrate the fit of $q$-Gaussians to PDFs. 

The $q$-indexes and their 95\% confidence intervals in each age bin were estimated using bootstrap resampling. Generally speaking, the bootstrap resampling  method consists of generating a large number of data sets, each with an equal amount of data randomly drawn from the original dataset and an estimator, which can be the mean or the median \cite{efron79,efron87,efron94,feigelson}. We choose the median because it is the measure less affected by possible outliers. The present study applied a similar procedure to that used by Silva, Soares \& de Freitas \cite{silva2014}, described as follows. First, we performed a set of 1000 bootstrap replications of the median $q$-index in the age bin extracted from the median value of the distribution of these bootstrapped samples, using the $q$-Gaussian with symmetric Tsallis distribution from Eq. (\ref{eq2}). Then, we ranked the bootstrapped medians from the lowest to the highest value and took the 25$^{th}$  and the 975$^{th}$  medians in the rank as the lower and upper limits of the confidence interval, respectively. The values of median $q$-indexes and their confidence intervals in each age bin and space velocity component are shown in Tables \ref{tab1} and \ref{tab2}.

To measure the random effects on the distribution of space velocities $U$, $V$ and $W$, we use the deviation of gaussianity, which is measured by the difference between the entropic index $q$ and its value that recovers the \textbf{BGS} statistical mechanics (i.e., $q = 1$). This parameter is defined as $q - 1$ and will be used as a measure of the degree of non-randomness in the process of heating the Galactic disc. Based on the works of Nordström {\it et al.} \cite{Nordstron} and Holmberg {\it et al.} \cite{Holmberg,Holmberg2}, $q-1$ should be close to $0$ (zero) for the $W$ component, regardless of age. Nevertheless, Figures \ref{fig1} and \ref{fig2} are in contrast with the conclusions of previous authors. These figures show the distributions of $q - 1$ as a function of stellar age for the F- and G-type stars, respectively. These distributions suggest that the velocities ($U, V, W$) are influenced by non-random processes, given that the values of $q - 1$ differ from zero.

In Figure \ref{fig1}, the values for components $U$, $V$ and $W$ for F-type stars have a more pronounced deviation for older ages than the solar one ($\sim$4.55 Gyr). For G-type stars (see Fig. \ref{fig2}), there is a strong indication that suggests that randomness does not govern any of the components of space velocity in any age range, except for the component $U$. Briefly speaking, this reveals that more massive stars, such as F-type stars, preserve randomness more efficiently than G-type stars. A quick search in the literature shows that a deviation of $q - 1$ of the order of 0.2 is large enough to infer a deviation from gaussianity \cite{Tsallis1,Tsallis3,deFreitas2}. Figures \ref{fig1} and \ref{fig2} clearly show that the mean value of $q - 1$ is of the order of 0.3.

In addition, this result shows that in both spectral types, the distributions of the $W$ component are out of the standard ``thermal equilibrium'', contrasting with the results mentioned by Nordström {\it et al.} \cite{Nordstron} and Holmberg {\it et al.} \cite{Holmberg}, \cite{Holmberg2}. When analysing Figures \ref{fig1} and \ref{fig2}, it is possible to deduce that all the components of the space velocity obey an anomalous diffusion as a function of age, where pure heating processes were not verified in most of the age groups analysed here.

For a more detailed follow-up analysis should be noted that several mechanisms have been proposed to explain the behaviour of $U$, $V$ and $W$ velocity distributions and, consequently, the kinematic heating of the Galactic disc \cite{Nordstron}. Such mechanisms are associated to non-axisymmetric structures of the Milky Way, e.g., past merger events \cite{helmi}, dynamical effects of the bar and spiral arms \cite{fux}, radial migration \cite{grand} and scattering by giant molecular clouds (GMCs) or star clusters \cite{aumer}.  These mechanisms are usually combined and difficult to isolate in practice. However, dynamical simulations suggest that the peanut-shaped bar had, at a subsequent epoch, an episode of buckling, which might affect the $W$ component and, therefore, the non-random nature of its distribution \cite{Nordstron,ma}.

\section{Concluding remarks}
Our work aimed to investigate how randomness can govern the space velocities triplet ($U, V, W$), especially the $W$ component. To this end, we used a sample of 6704 stars composed of F- and G- type field stars in the solar neighbourhood, extracted from the GCS. The samples were segregated by age interval (step size of 1 Gyr) and the medians of entropic index $q$, as well as the 95\% confidence intervals of these medians, were calculated using the non-parametric bootstrap resampling method.

First, it is notable that there is a smooth growth for the $U$ and $W$ components of F-type stars around the solar age. However, the $V$ component related to the tangential component in the direction of Galactic rotation shows an oscillating behaviour in all age ranges. For G-type stars, the increase in dispersion in the velocity components ($U, V, W$) occurs smoothly at all ages.

We also measured deviations of randomness, denoted by parameter $q - 1$, during almost the entire life of the stars in the Milky Way ($\sim$10 Gyr). As a result, we observed that there are heating mechanisms with a non-random nature that affect space velocities ($U, V, W$), disagreeing with results in the literature that point to a totally pure, or random, heating for the $W$ component of velocity. It is notable that for F-type stars, only a region close to 2.5 Gyr seems to be dominated by randomness for the three components. These results are very important, as they reveal that the distribution of the $W$ component in particular does not ``heat up'' at random, as has been predicted by previous works that attributed this evolution to a mechanism totally governed by randomness. Unfortunately, our results were not able to reveal which mechanisms act on the $W$ component. Nevertheless, we can affirm that due to the degree of interaction that exists between the components of space velocity, there must certainly be different sources that act by modifying the degree of mixing of the stars in the Galactic disc.

Finally, it is worth mentioning that the entropic index $q$ can also be used to investigate the possible sources for the disc heating rate and, therefore, a deeper analysis may have origins in the metallicity and galactic location. In addition, it would be interesting to investigate this issue using more robust data from big surveys such as the GAIA and RAVE programmes \cite{stein}. This issue will be addressed in a forthcoming communication.

\acknowledgments
DBdeF acknowledges financial support from the Brazilian agency CNPq-PQ2 (Grant No. 311578/2018-7). Research activities of STELLAR TEAM of Federal University of Cear\'a are supported by continuous grants from the Brazilian agency CNPq. The authors would like to dedicate this paper to all the victims of the COVID-19 pandemic around the world.

\end{document}